\documentclass{cupconf}

  \checkfont{eurm10}
  \iffontfound
    \IfFileExists{upmath.sty}
      {\typeout{^^JFound AMS Euler Roman fonts on the system,
                   using the 'upmath' package.^^J}%
       \usepackage{upmath}}
      {\typeout{^^JFound AMS Euler Roman fonts on the system, but you
                   dont seem to have the}%
       \typeout{'upmath' package installed. cupconf.cls can take advantage
                 of these fonts,^^Jif you use 'upmath' package.^^J}%
      }
  \else
  \fi

  \checkfont{msam10}
  \iffontfound
    \IfFileExists{amssymb.sty}
      {\typeout{^^JFound AMS Symbol fonts on the system, using the
                'amssymb' package.^^J}%
       \usepackage{amssymb}%
         \let\leq=\leqslant
         
      }{}
  \fi

  \IfFileExists{amsbsy.sty}
    {\typeout{^^JFound the 'amsbsy' package on the system, using it.^^J}%
     \usepackage{amsbsy}}
    {}

\newsavebox{\astrutbox}
\sbox{\astrutbox}{\rule[-5pt]{0pt}{20pt}}

\input psfig.sty
\title[Wolf-Rayet Winds]{Metallicity Dependent Wolf-Rayet Winds}

\author[P. A. Crowther]{P\ls A\ls U\ls L\ns A.\ns C\ls R\ls O\ls W\ls T\ls H\ls E\ls R$^1$}

\affiliation{$^1$Department of Physics \& Astronomy, University of Sheffield,
Hicks Building, Hounsfield Road, Sheffield, S3 7RH, UK}

\pubyear{2006}
\volume{XXX}
\pagerange{XXX}
\date{?? and in revised form ??}
\begin{document}

\maketitle

\begin{abstract}
Observational and theoretical evidence in support of metallicity dependent winds
for Wolf-Rayet stars is considered. Well known differences in Wolf-Rayet subtype 
distributions in the Milky Way, LMC and SMC may be attributed to the sensitivity
of subtypes to wind density. Implications for Wolf-Rayet stars at low metallicity 
include a hardening of ionizing flux distributions, an increased WR population
due to reduced optical line fluxes, plus support for  the role  of single WR stars as 
Gamma Ray Burst progenitors.
\end{abstract}

\firstsection 
\section{Introduction}

Wolf-Rayet (WR) stars represent the final phase in the evolution
of very massive stars prior to core-collapse, in which the H-rich
envelope has been stripped away via either stellar winds or close
binary evolution, revealing products of H-burning (WN sequence) or
He-burning (WC sequence) at their surfaces, i.e. He, N or C, O (Crowther
2007). 

\begin{figure}
\centerline{\psfig{file=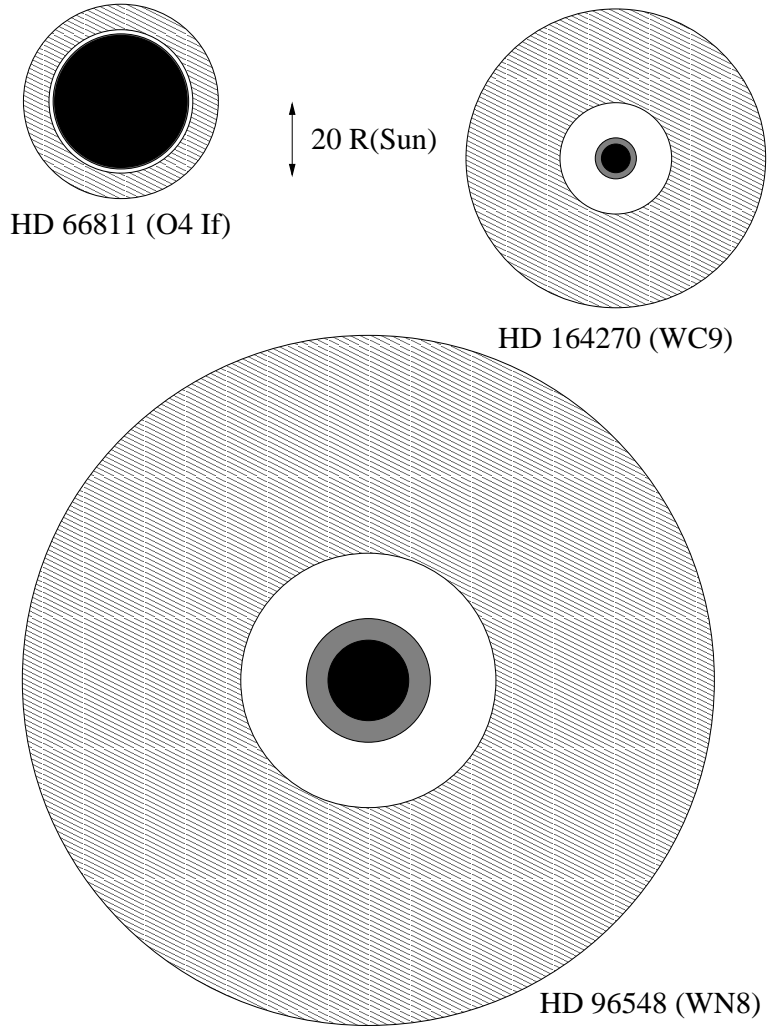,width=4.0in}}
\caption{Comparisons between stellar radii at Rosseland
optical depths of 20 (= $R_{\ast}$, black) and 2/3 (= $R_{2/3}$, grey)
for HD~66811 (O4\,If), HD~96548 (WN8) and HD~164270 (WC9), shown to
scale, together with the wind region corresponding to the primary
optical wind line forming region, $10^{11} \leq n_{e} \leq 10^{12}$
cm$^{-3}$ (hatched) in each case, illustrating the highly extended
winds of WR stars with respect to O stars (Crowther 2007).}\label{WRross}
\end{figure}

WR stellar winds are significantly denser than O stars, as illustrated
in Fig.~\ref{WRross}, so their visual spectra are dominated by broad emission lines, 
notably HeII $\lambda$4686 (WN stars) and CIII $\lambda$4647-51, CIII $\lambda$5696, CIV 
$\lambda$5801-12 (WC 
stars). The spectroscopic signature of WR stars may be seen individually
in Local Group galaxies (e.g. Massey \& Johnson 1998), within knots in 
local star forming galaxies (e.g. Hadfield \& Crowther 2006)
and in the average rest frame UV spectrum of Lyman Break Galaxies
(Shapley et al. 2003).

In the case of a single massive star, the strength of stellar winds during
the main sequence and blue supergiant phase scales with the metallicity
(Vink et al. 2001). Consequently, one expects a higher threshold for the
formation of WR stars at lower metallicity, and indeed the SMC shows a
decreased number of WR to O stars than in the Solar Neighbourhood.  
Alternatively, the H-rich envelope may be removed during the Roche lobe
overflow phase of close binary evolution, a process which is not expected
to depend upon metallicity.


WR stars represent the prime candidates for Type Ib/c core-collapse
supernovae and long, soft Gamma Ray Bursts (GRBs).  This is due to their
immediate progenitors being associated with young massive stellar
populations, compact in nature and deficient in either hydrogen (Type Ib)
or both hydrogen and helium (Type Ic).  For the case of GRBs, a number of
which have been associated with Type Ic hypernovae (Galama et al. 1998;
Hjorth et al. 2003), a rapidly rotating core is a requirement for the
collapsar scenario in which the newly formed black hole accretes via an
accretion disk (MacFadyen \& Woosley 1999).  Indeed, WR populations have
been observed within local GRB host galaxies (Hammer et al. 2006).

At solar metallicity, single star models predict that the core is spun
down either during the red supergiant (via a magnetic dynamo) or
Wolf-Rayet (via mass-loss)  phases. The tendency of GRBs to originate from
metal-poor environments (e.g. Stanek et al. 2006) suggests that stellar
winds from single stars play a role in their origin since Roche lobe
overflow in a close binary evolution would not be expected to show a
strong metallicity dependence.

In this article, evidence in favour of a metallicity dependence for WR
stars is presented, of application to the observed WR subtype distribution
in Local Group galaxies, plus properties of WR stars at low metallicity
including their role as GRB progenitors.

\begin{figure}
\centerline{\psfig{file=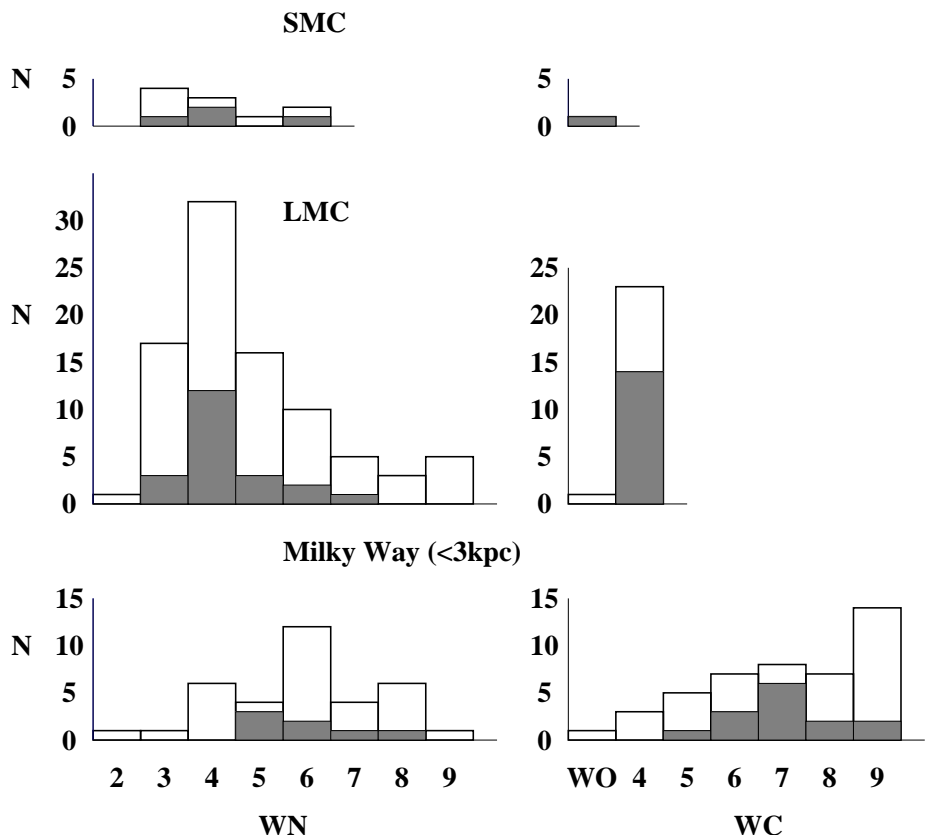,width=4.8in}}
\caption{Subtype distribution of Milky Way ($<$3kpc),
LMC and SMC WR stars, in which known binaries are
shaded (Crowther 2007).}\label{wrpop}
\end{figure}
\section{WR subtype distribution}

Historically, the wind properties of WR stars have been assumed to be
metallicity independent (Langer 1989), yet there is a well known
observational trend to earlier, higher ionization, WN and WC subtypes at
low metallicity as illustrated in Fig.~\ref{wrpop}, whose origin is yet to
be established.

\begin{figure}
\centerline{\psfig{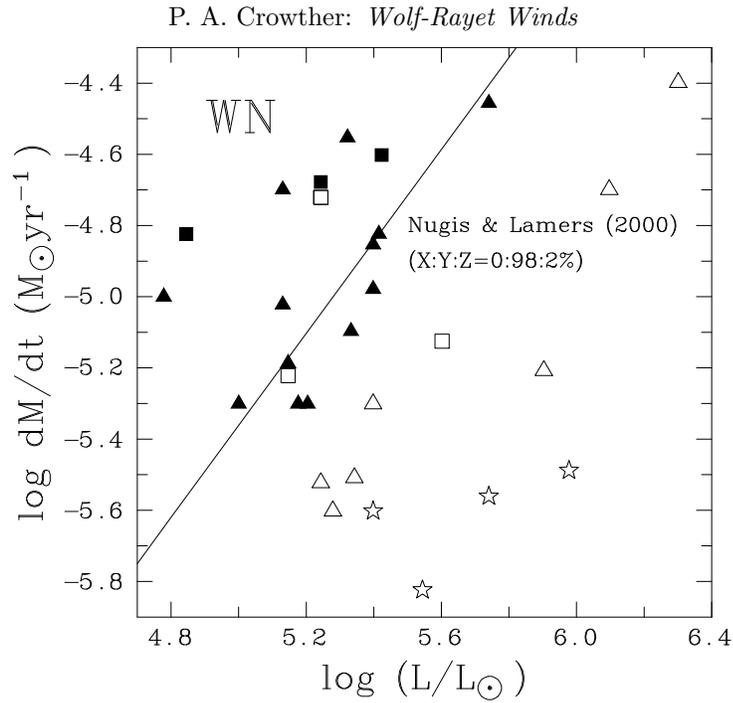}}
\caption{Mass-loss rates for WN stars in the Galaxy (squares), 
LMC (triangles) and SMC (stars) revealing a wide spread in wind densities for
WN stars, for which stars without hydrogen (filled symbols) possess
stronger winds. The solid line is from eqn~22 of Nugis \& Lamers (2000).}
\label{wn_mdot}
\end{figure}

\begin{figure}
\centerline{\psfig{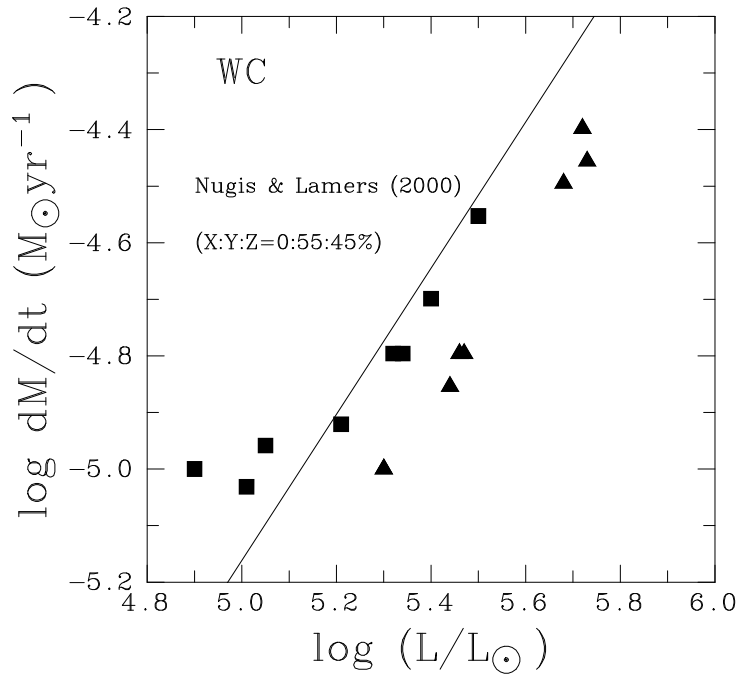}}
\caption{Mass-loss rates for WC and WO stars in 
in the Galaxy (squares) and LMC (triangles). The
solid line is from eqn~22 of Nugis \& Lamers (2000).}
\label{wc_mdot}
\end{figure}

Mass-loss rates for WN stars in the Milky Way and LMC show a very large
scatter. The presence of hydrogen in some WN stars further complicates the
picture since WR winds are denser if H is absent (Nugis \& Lamers 2000).
This is illustrated in Fig.~\ref{wn_mdot}, which reveals that the wind
strengths of (H-rich) WN winds in the SMC  are lower than corresponding 
H-rich stars in  the LMC and Milky Way (Crowther 2006). Fig.~\ref{wc_mdot} 
shows that the situation is rather clearer for WC stars, for which LMC 
stars reveal  $\sim$0.2 dex lower mass-loss rates than  Milky Way 
counterparts (Crowther  et al. 2002)
 
The observed trend to earlier subtypes in the LMC (Fig.~\ref{wrpop}) was
believed to originate from a difference in carbon abundances relative to
Galactic WC stars (Smith \& Maeder 1991), yet quantitative analysis
reveals similar carbon abundances (Koesterke \& Hamann 1995; Crowther et
al. 2002).

\begin{figure}
\centerline{\psfig{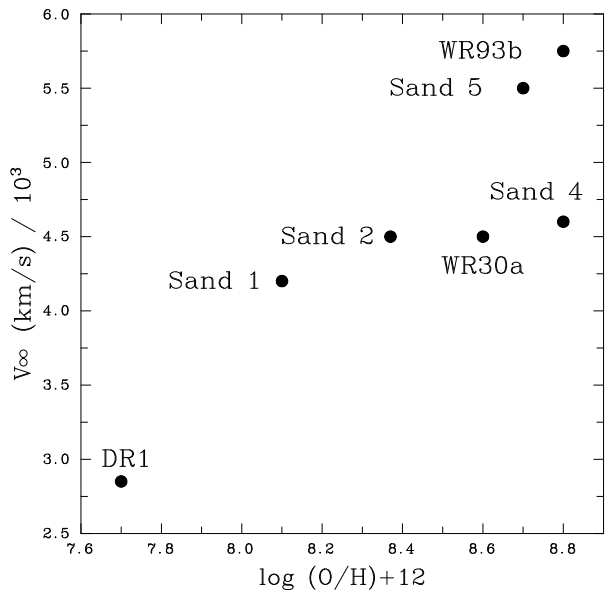}}
\caption{Wind velocities for WO stars as a function of metallicity (Crowther \&
Hadfield 2006).}\label{wo}
\end{figure}

Theoretically, Nugis \& Lamers (2002) argued that the iron opacity peak
was the origin of the wind driving in WR stars, which Gr\"{a}fener \&
Hamann (2005) supported via an hydrodynamic model for an early-type WC
star in which lines of Fe IX-XVII deep in the atmosphere provided the
necessary radiative driving. Vink \& de Koter (2005) applied a Monte Carlo
approach to investigate the metallicity dependence for cool WN and WC
stars revealing $\dot{M} \propto Z^{\alpha}$ where $\alpha$=0.86 for WN
stars and $\alpha$=0.66 for WC stars for 0.1 $\leq Z \leq 1 Z_{\odot}$.
The weaker WC dependence originates from a decreasing Fe content and
constant C and O content at low metallicity. Empirical results for the
Solar neighbourhood, LMC and SMC presented in 
Figs.~\ref{wn_mdot}--\ref{wc_mdot} are broadly consistent with theoretical 
predictions, although detailed studies of  individual WR 
stars within galaxies broader range in metallicity would provide stronger
constraints. Theoretical wind models also
predict smaller wind velocities at lower metallicity, as is 
observed for  WO
stars, which are presented in Fig.~\ref{wo} (Crowther \& Hadfield 2006).

\begin{figure}
\centerline{\psfig{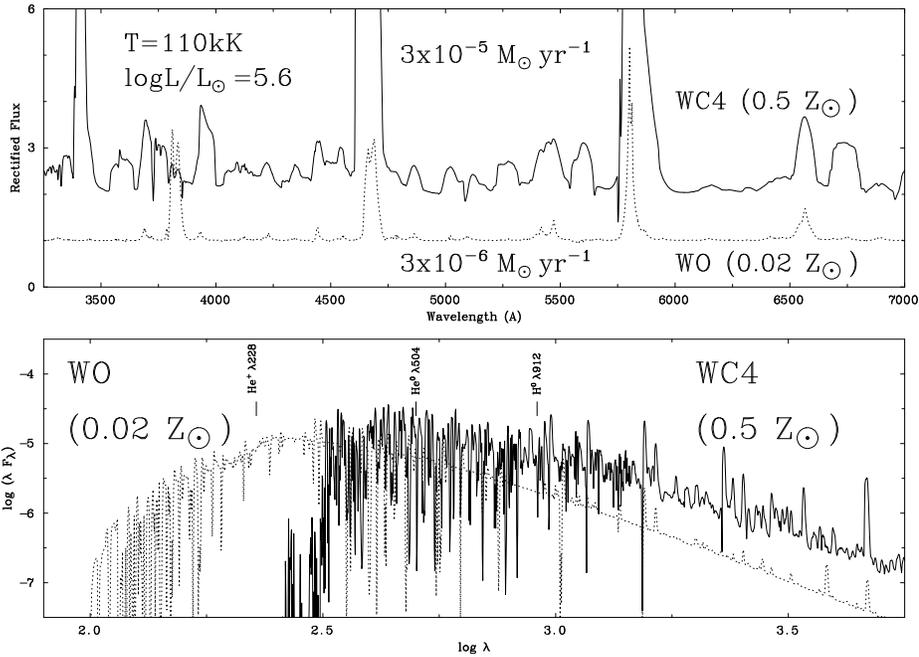}}
\caption{Comparison between theoretical stellar atmosphere models
which differ solely in wind density (factor of 10), revealing 
an earlier spectral type (WC4 $\rightarrow$ WO) and harder ionizing
flux distribution at low metallicity, adapted from Crowther \& Hadfield 
(2006)}\label{wc_wo}
\end{figure}

The impact of a metallicity dependence for WR winds upon spectral
types is as follows. At high metallicity, recombination from high
to low ions (early to late subtypes) is very effective in very
dense winds, whilst the opposite is true for low metallicity, low
density winds. The situation is illustrated in the upper 
panel of Fig.~\ref{wc_wo}, where we present synthetic WC spectra obtained
from   identical models except that their
wind densities differ by a factor of 10, and the weak wind model is 
assumed to be extremely Fe-poor (adapted from Crowther \& Hadfield 2006). 
The high wind density case has a WC4 
spectral type whilst the low wind density case has an earlier WO subtype. 
Crowther et al. (2002) noted that a further increase in wind density 
by a factor of 2 predicts a WC7 subtype. Stellar temperatures 
further complicates this picture, such that the spectral type of a WR star 
results from a subtle combination of  ionization and wind density, 
in contrast with normal stars.

\section{WR populations at low metallicity}

The effect of reduced WR wind densities at low metallicity on WR
populations is as follows. WR optical recombination lines will (i) 
decrease in equivalent width, since their strength scales with the square 
of the  density, and (ii) decrease in line flux, since the lower
wind strength will reduce the line blanketing, resulting in an increased
extreme UV continuum strength at the expense of the UV and optical.
The equivalent widths of optical emission  lines in SMC WN stars are well
known to be lower than Milky Way and LMC counterparts (Conti et al. 1989).
To date, the standard approach for the determination of unresolved WR
populations in external galaxies has been to assume metallicity 
independent WR line fluxes --  obtained for Milky Way and LMC stars (Schaerer 
\& Vacca 1998) -- regardless of whether the host galaxy is metal-rich
(Mrk 309, Schaerer et al. 2000) or metal-poor (I~Zw\,18, Izotov et al. 1997).

Ideally, one would wish to use WR template stars 
appropriate to the metallicity of the galaxy under consideration. 
Unfortunately, this is only feasible for the LMC, SMC and Solar neighbourhood, since
it is challenging to isolate individual WR stars from ground based 
observations in more distant galaxies, which span a larger spread in
metallicity. Line luminosities for optical emission lines in LMC and SMC WR stars 
are  compared in Table~\ref{flux}, illustrating significantly lower (factor of 5--6)
luminosities for the lower metallicity of the SMC.

\begin{table}
\begin{center}
\begin{tabular}{lcrcr}
    & \multicolumn{2}{c}{LMC }     & \multicolumn{2}{c}{SMC } \\
    & $\log L$ (erg s$^{-1}$)   &  N   & $\log L$ (erg s$^{-1}$) & N  \\[3pt]
WN2--4 & 35.92                  & 36   & 35.23     & 8  \\ 
WN5--6 & 36.24                  & 15   & 35.63     & 2   \\
WN7--9 & 35.86                  & 9    \\
\end{tabular}
\end{center}
\caption{Mean HeII $\lambda$4686 line luminosities  for Magellanic Cloud WN stars
including known binaries (Crowther \& Hadfield 2006)}\label{flux}
\end{table}

Reduced WR line fluxes are also predicted for WR atmospheric models
at low metallicity if one follows the metallicity dependence from Vink \& 
de Koter (2005), such that WR populations inferred from
Schaerer \& Vacca (1998) at low metallicity may underestimate actual
populations by an order of magnitude. This is potentially problematic
for single star evolutionary models at very low
metallicities ($\sim 1/50 Z_{\odot}$) since the WR populations inferred 
for I~Zw\,18 and SBS0335-052E using Milky Way line fluxes compare well
with evolutionary models (e.g. Izotov et al. 1997; Papaderos et al. 2006). 
If WR populations
are in fact a factor of $\sim$10 larger, similar to that of the SMC, close 
binary evolution would represent the most likely origin for such large
WR populations.

\begin{figure}
\begin{tabular}{cc}
\multicolumn{2}{c}{\psfig{file=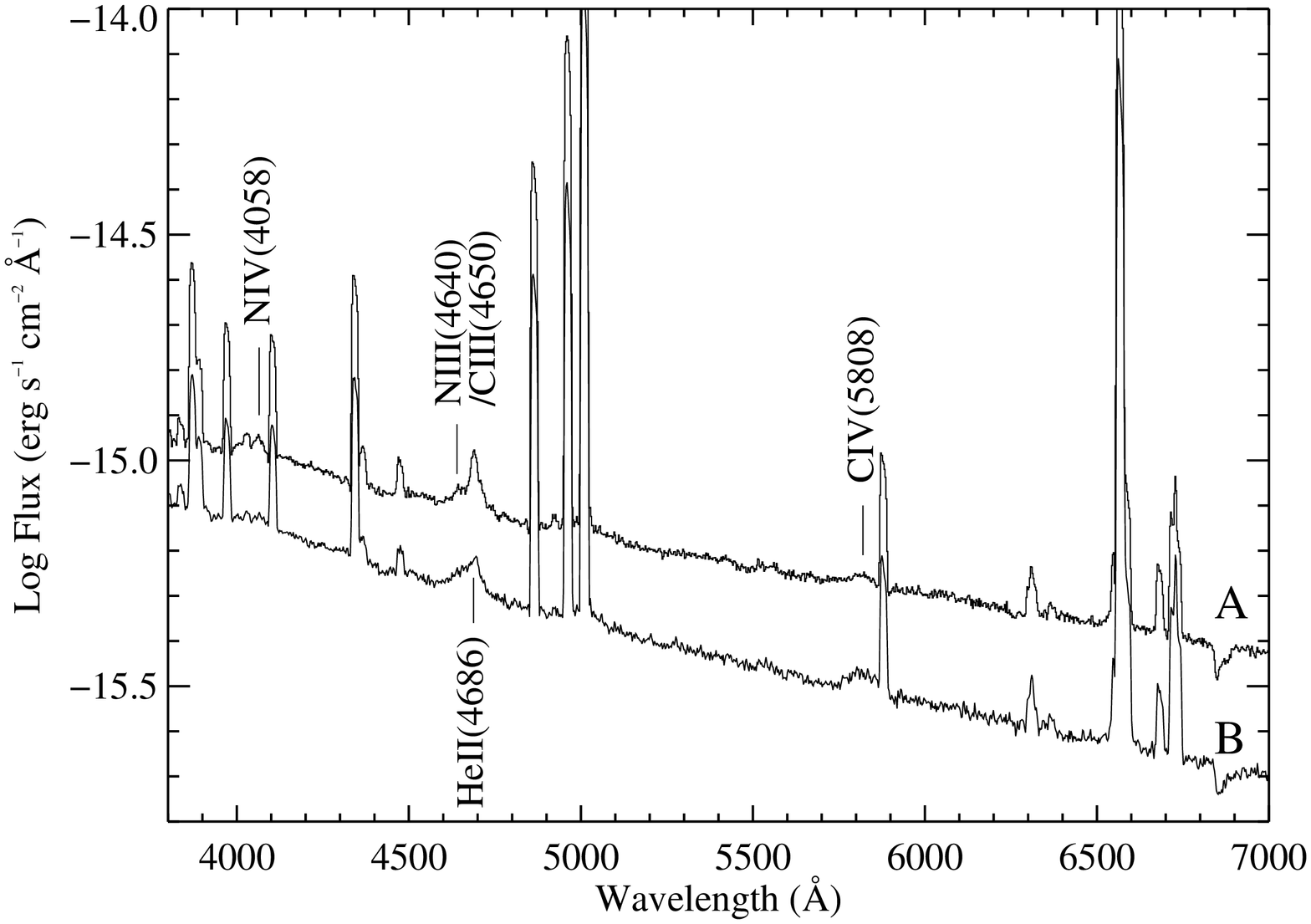,width=4.5in,angle=90}} \\
\psfig{file=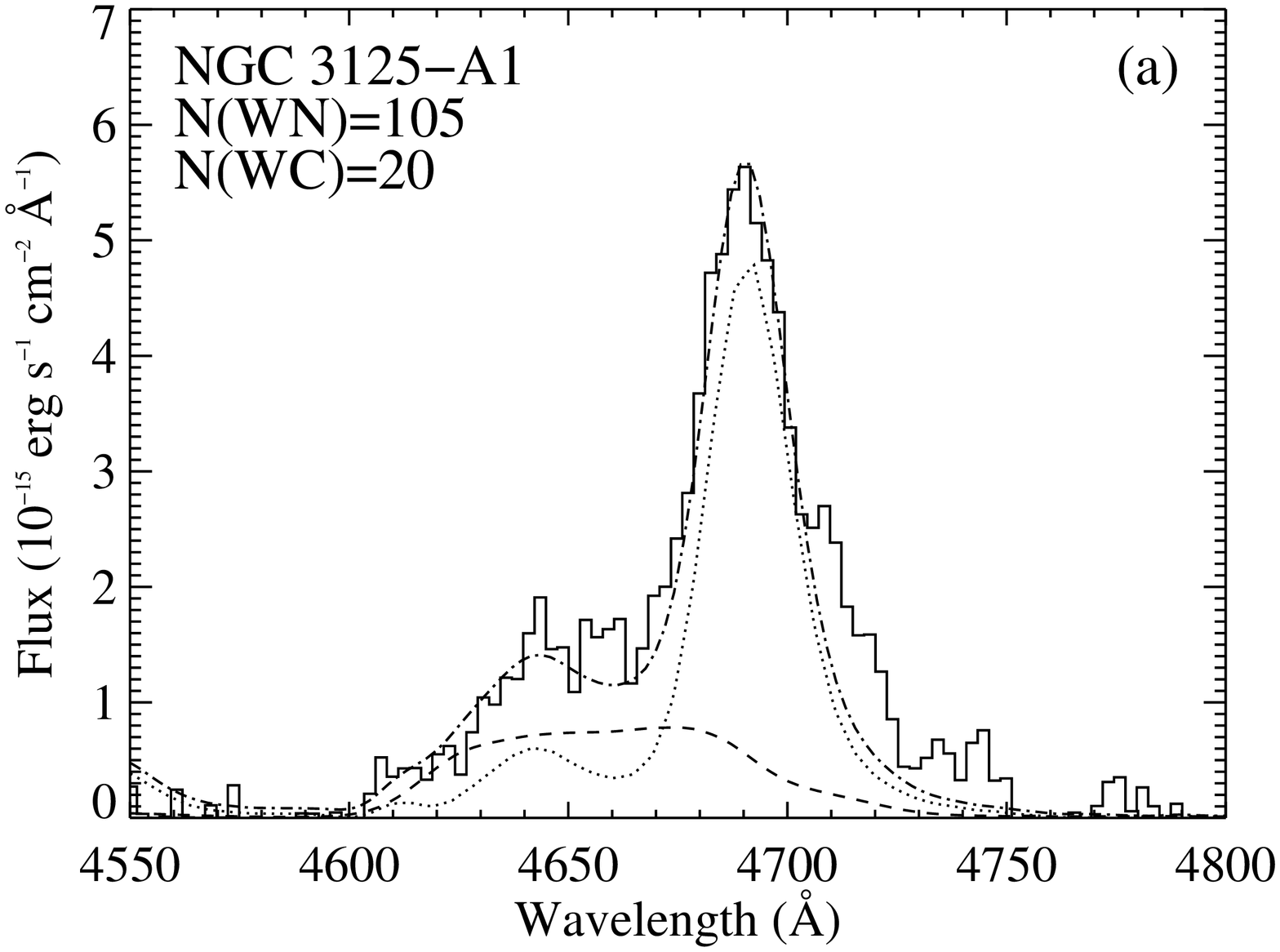,width=2.4in,angle=90} &
\psfig{file=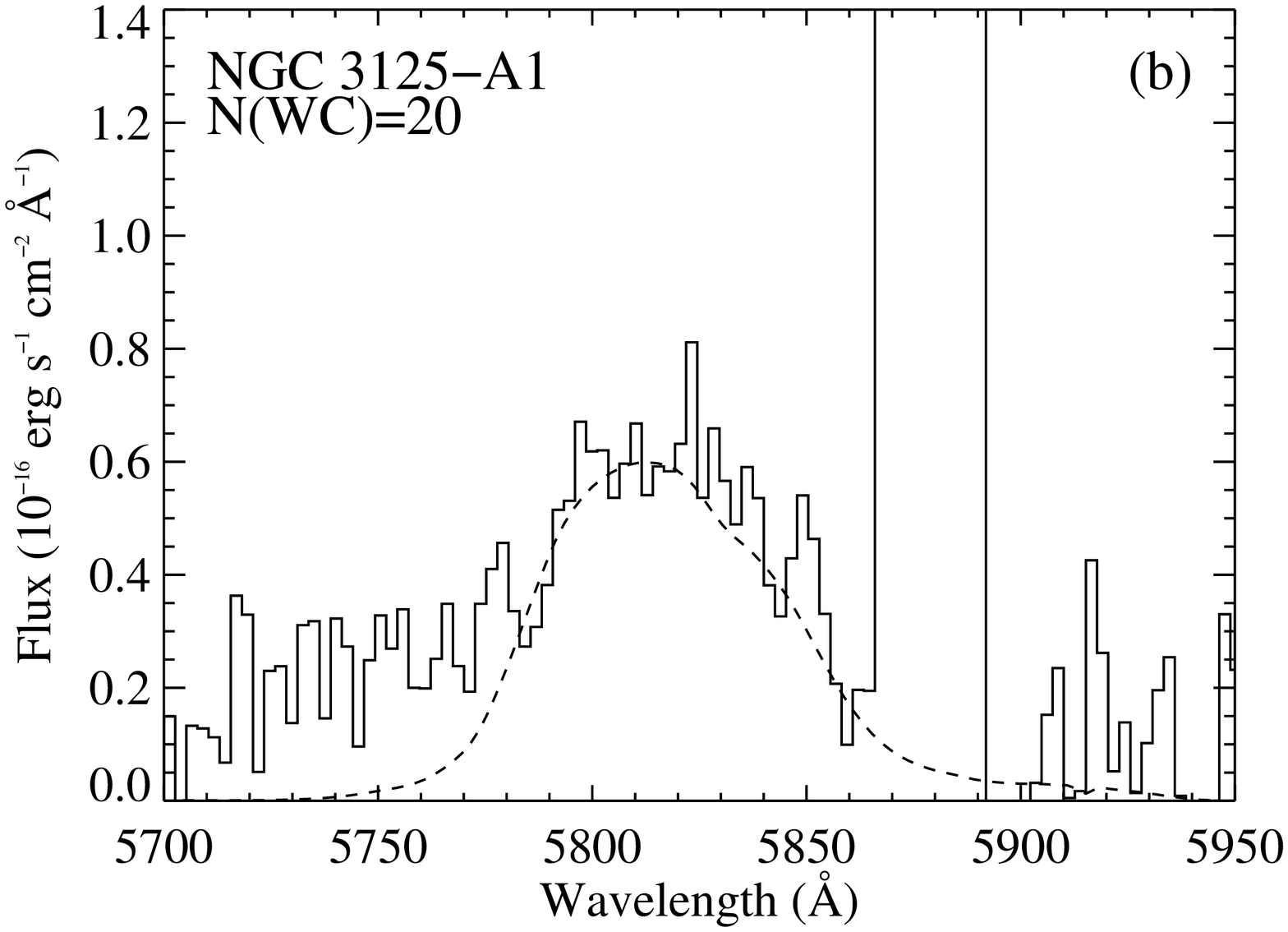,width=2.4in,angle=90} \\
\end{tabular}
\caption{Top panel:
Optical spectroscopy of knots A and B within the LMC metallicity starburst 
galaxy NGC~3125, indicating the CIII 4650/HeII 4686 (blue) and CIV 5801-12 (yellow) 
WR bumps. Lower panels: Fit to the blue and yellow bumps for cluster A1 using LMC template
WN (dotted lines) and WC (dashed lines) stars (Hadfield \& Crowther 2006)}\label{ngc3125}
\end{figure}

For Magellanic Cloud metallicity starburst galaxies, one may employ appropriate template
spectra (e.g. Crowther \& Hadfield 2006) to reproduce WR features, as shown
 in the upper panel of Fig.~\ref{ngc3125} for the starburst galaxy NGC~3125. 
Indeed, consistent fits to the
blue and yellow WR bumps may be achieved for the A1 cluster within NGC~3125 using LMC
template WR stars, as shown in the lower panels of 
Fig.~\ref{ngc3125} (Hadfield \& Crowther 2006).

\section{Ionizing fluxes and GRB progenitors}

Schmutz et al. (1992) demonstrated that the ionizing fluxes from WR stars
soften as wind density increases. Consequently, a metallicity dependence
for WR wind strengths implies that WR ionizing flux distributions soften
at increased metallicity, as demonstrated by Smith et al. (2002). Indeed,
relatively soft ionizing fluxes are observed in the super-Solar 
metallicity WR starburst galaxy NGC~3049 (Gonzalez Delgado et al. 2002).

At low metallicities, one anticipates a combination of weak UV and optical
spectral lines from WR stars (i.e. weak stellar HeII $\lambda$4686) but
very strong H and He Lyman continua (i.e. strong nebular HeII
$\lambda$4686), as is indicated in the lower panel of Fig.~\ref{wc_wo}. 
Indeed, low metallicity
star forming galaxies display strong nebular HeII $\lambda$4686, although
shocks from supernovae remnants may also contribute to nebular emission.

The typically environment of nearby ($z < 0.25$) 
long duration GRBs is unusually 
metal-poor, as emphasized by
Stanek et al. (2006) with respect to star forming galaxies from
the Sloan Digital Sky Survey.
Reduced WR mass-loss rates at low metallicity will lead to reduced 
densities in the immediate environment of GRBs with respect to typical
WR stars, as is observed (Chevalier et al. 2004). In addition, massive
single stars undergoing homogeneous evolution in which WR mass-loss rates 
are low may maintain their rapidly spinning cores through to core-collapse
(Yoon \& Langer 2005; Langer \& Norman 2006).

\section{Summary}

Observational and theoretical evidence supports reduced wind densities and
velocities for low metallicity WR stars, which addresses the relative WR 
subtype distribution in the Milky Way and Magellanic Clouds, plus the 
reduced WR line strengths in the SMC with regard to the Galaxy and LMC.
The primary impact at low metallicity is as follows; (a) an increased
WR population due to lower line fluxes from individual stars, of 
particular relevance to I~Zw\,18 and SBS0335-052E; (b) harder ionizing
fluxes from WR stars, potentially responsible for the strong nebular
HeII $\lambda$4686 seen in low metallicity HII galaxies; (c) responsible for the
reduced density of GRB environments with respect to Solar metallicity 
WR counterparts.

Finally, a metallicity dependence for WR winds may help to reconcile the
relative number of WN to WC stars observed in surveys (e.g. Massey \& Johnson
1998) with evolutionary predictions. Evolutionary models
for which rotational mixing is included yet metallicity dependent WR winds are
not (Meynet \& Maeder 2003) fail to predict the high N(WC)/N(WN) ratio observed
at high metallicities (Hadfield et al. 2005), whilst models which account for
the Vink \& de Koter (2005) WR wind dependence compare much more favourably with
observations (Eldridge \& Vink 2006), in spite of the neglect of rotational mixing.

\begin{acknowledgements}
Many thanks to Lucy Hadfield, with whom the majority of the results presented
here were obtained. PAC acknowledges financial support from the Royal Society.
\end{acknowledgements}

\end{document}